\documentclass[aps,prl,showpacs,preprint]{revtex4-1}
\usepackage{epsfig}
\usepackage{graphics}
\usepackage{caption}
\usepackage{subcaption}
\usepackage{graphicx}
\usepackage{epstopdf}
\DeclareGraphicsExtensions{,.pdf,.png,.jpg,.gif}

\begin{document}

\title{Classical dynamics emerging from quantum dynamics in macroscopic bodies, a note with a simple example.}

\author{V\'{\i}ctor Romero-Roch\'in} 

\email{romero@fisica.unam.mx}

\affiliation{Instituto de F\'{\i}sica, Universidad
Nacional Aut\'onoma de M\'exico. \\ Apartado Postal 20-364, 01000 M\'exico D.
F. Mexico.}


\date{\today}

\begin{abstract}
Using very general and well established ideas of the statistical physics of macroscopic bodies, that is, of those composed of many degrees of freedom, we show how classical behavior of the center of mass motion arises from a fully quantum mechanical description of the dynamics of the whole body. We do not attempt to provide a rigorous proof of the latter statement, but rather, we show or, at least, indicate the hypotheses needed to obtain the purported result. Moreover, we neither attempt to deal with the ``most general" physical situation and, instead, we concentrate on a stylized model of a small solid, yet macroscopic, that we shall call a ``little stone". The main hypothesis is that a macroscopic body can be decomposed into several smaller pieces, still macroscopic, that become statistically independent 
due to the short-range interaction nature of their constituent  atoms. The ensuing main result is that the quantum distributions of extensive variables of the body become sharply-peaked. The center of mass variables are of this type and hence their dynamics is essentially classical. We point out the crucial role played by the external potential, in which the motion occurs, as the macroscopic agent that executes the ``measurement" process of the center of mass.

\end{abstract}

\maketitle

\section{Introduction}

Deriving classical behavior from a quantum one is a question that we all, practitioners of physics, have faced at some point of our careers. This is because we believe that the world is quantum and that classical behavior is an approximation. We are also aware that classical mechanics should apply to macroscopic bodies, such as balls, stones, houses, rockets, planets, et cetera. We know very well that at the atomic or molecular level, the dynamics must be described with quantum mechanics \cite{Cohen}, although there are instances in which, even at that level, classical mechanics works fine for the description of many physical properties. Therefore, there is always the question of when and how classical mechanics ``emerges" from an otherwise quantum description. While there are related questions to the one posed here, that go under the names of ``decoherence" and the ``measurement" problem\cite{Zurek}, our goal is not to dwell into concepts like those, but rather to simply make use of well established ``facts" of the discipline of statistical physics\cite{LL}, when explicitly applied to bodies composed of a very large number of atoms. True, we borrow those concepts, well defined in thermal equilibrium, and simply extend them to non-equilibium situations and, hence, we may leave dissatisfied more than one reader. But even if we cannot prove the main requirements, we believe that, at least, we point out what those requirements are, in order to obtain the classical limiting behavior of macroscopic bodies.

As we shall elaborate in some detail below, the main hypothesis is that macroscopic bodies can be decomposed into smaller, but still macroscopic parts that can be considered as {\it statistically independent}. The physical reason behind this statement is the fact that atoms and molecules in macroscopic systems effectively interact via short-range potentials. As it can be shown\cite{LL}, ``short-range" means that the potentials decay as $r^{-\alpha}$, with $\alpha > 3$ and $r$ the distance between the interacting atoms. Once one accepts this hypothesis, it follows that the (probability) distributions of {\it extensive} quantities are sharply-peaked, namely, that their fluctuations are $1/\sqrt{N}$ smaller than their mean values, with $N$ the number of involved degrees of freedom. Hence, if $N \sim 10^{20}$ the values of extensive variables become accurately and uniquely determined, at any time. These statements are the basis of statistical physics, as clearly described in any textbook, such as Landau and Lifshitz {\it Statistical Physics}\cite{LL}. What we will expose here is that the momentum and position of the center of mass of a (solid) macroscopic body are extensive variables and, therefore, their distributions are sharply peaked. The obtained averages, or means, of those variables then obey classical mechanics, with ``dissipative" terms that simply reflect the interaction of the center of mass with the very large number of {\it relative} degrees of freedom of the body. The latter degrees of freedom do not need to obey classical mechanics; indeed, if the temperature of the ``little stone" is low enough, the relative degrees of freedom must be treated quantum mechanically. We insist that we shall focus on the model of a little solid stone because we want to ``derive" Newton equations for its center of mass. If we dealt with a fluid, we would then obtain Navier-Stokes-like equations, a much technically harder endeavor, but that in any case could be realized with the same ideas here discussed; see also Ref. \cite{Zurek}.  An additional, and very important requirement to make possible the appearance of classical behavior, is the presence of an {\it external} anharmonic ``macroscopic" potential. That is, for the different macroscopic subsystems of the little stone to be able to continuously ``measure" the center of mass position and momentum, the relative and center of mass degrees of freedom must be coupled and, as we shall show, this can only be achieved if the external potential has anharmonic terms. Otherwise, the center of mass is always decoupled from the macroscopic number of relative coordinates, and one can no longer appeal to the statistical independence of its parts.
%
It is interesting, therefore, to realize that the presence of macroscopic agents, namely, the external bodies that produce the external field, are ultimately responsible for the ``measurement" and, in common but dangerous language, one may say that the external agent and the many relative degrees of freedom ``collapse" the quantum evolution of the center of mass into a single, causal, classical evolution. This is not so unfamiliar. In equilibrium systems, the presence of a macroscopic confining potential, such a box of rigid walls, is also partly responsible for the sharply-peaked values of their internal energy.

%
%
%

Let us define a simple model of a little stone as a system of $N$ identical atoms interacting pairwise in the presence of an external potential,
\begin{equation}
\hat H = \sum_{i=1}^N \frac{\hat{\vec p}_i^2}{2m} + \sum_{i<j} u(|\hat {\vec r}_i - \hat {\vec r}_j |) + \sum_{j = 1}^N V_{\rm ext}(\hat {\vec r}_j) . \label{H0}
\end{equation}
The ``hats" on all variables is to insist that those are quantum operators. It is assumed that $N \gg 1$. Since we are assuming the temperature of the system is low enough to consider it as being solid, we should perhaps include more complicated interatomic potentials. However, the point is to make explicit that the interatomic potentials depend on relative coordinates, $\hat {\vec r}_i - \hat {\vec r}_j$. The interatomic potential is assumed to be short-range, namely that decays faster than $r^{-3}$ and their interaction range $\sigma$ is of atomic length scales, say a few Angstroms. The external field is ``macroscopic" in the sense that its spatial variations are in length scales much larger than atomic ones; in other words, the quantization of a single atom in the potential $V_{\rm ext}(\hat {\vec r})$ must give rise to energy levels that are separated in a scale much smaller than in any atomic energy contribution. It should be produced by other macroscopic bodies that do not feel the reaction of the stone, such as large coils, magnets, lasers or gravity. It is further assumed that all atoms feel the same external potential and, for simplicity, it should have a minimum. As it will be explicitly used, it must contain anharmonic contributions.

There should not be any questioning into considering the system as ``closed", namely, that its only interaction with the ``environment" is through the external potential. Under these conditions, the state of the system can always be considered as given by a density matrix $\hat \rho(t)$, at any time, that obeys Schr\"odinger equation,
\begin{equation}
i\hbar \frac{\partial}{\partial t} \hat \rho(t) = \left[\hat H, \hat \rho(t)\right] \label{denmat}
\end{equation}
provided a given initial condition $\hat \rho(0)$. As it will be mentioned at the end, for systems  with $N \gg 1$, in real life there is not much arbitrariness in the specification of the initial state. Therefore, although not essential, we shall assume that the initial state has a well defined total energy, within the best experimental accuracy, that is, $E_T = \langle \hat H \rangle$ with variance $\delta E^2 = \langle \hat H^2 \rangle - \langle \hat H \rangle^2$, such that $\delta E \ll E_T$. The average or expectation value of any physical, Hermitian variable of the system, $\hat A = \hat A(\hat{\vec r}_1, \dots, \hat{\vec r}_N; \hat{\vec p}_1, \dots, \hat{\vec p}_N)$, at any time $t$ is given by,
\begin{eqnarray}
\langle \hat A (t) \rangle &= & {\rm Tr} \> \left( \hat \rho(t) \hat A \right) \nonumber \\
&=&  {\rm Tr} \> \left( \hat \rho(0) \hat A(t) \right).
\end{eqnarray}

Because the interatomic interactions are short-range, the little solid stone can be thought to be composed of ${\cal N}$ macroscopic pieces, or subsystems, that interact with each other via the atoms at their borders only. The main assumption of this work is that those subsystems become {\it statistically independent}. This is of course true only approximately, yet this is the basic assumption of statistical physics.  Let us denote by $\hat A_s$, with $s = 1, 2, 3, \dots, {\cal N}$ variables that pertain to each subsystem. This means that $\hat A_s$ essentially depends only on degrees of freedom of the $s$-th subsystem. Using the full density matrix $\hat \rho$, one can construct the distribution of values of any variable $\hat A_s$ by calculating all the moments of $\hat A$ at any given state, namely, calculating $\langle \hat A_s^n \rangle$ for all $n = 1, 2, ... \infty$. This means that we can, in principle, calculate ${\cal W}(A_s)$, the distribution of values of $\hat A_s$ in the given state. Statistical independence means that the distribution of all variables  $\hat A_s$, with $s = 1, 2, 3, \dots, {\cal N}$ is,
\begin{equation}
W(A_1, A_2, \dots, A_{\cal N}) \approx  {\cal W}(A_1){\cal W}(A_2) \dots {\cal W}(A_{\cal N}) .
\end{equation}
While this is true in equilibrium, we extend it to ``arbitrary" states, realizable in typical situations.

Of essential importance are the variables which are {\it extensive} or additive. That is, those that obey,
\begin{equation}
\hat A_{1-M} = \hat A_1 + \hat A_2 + \dots + \hat A_M
\end{equation}
where $\hat A_{1-M}$ is the value of the same variable, but for all the subsystems 1 to $M$ together. These type of variables commute with each other. The main property of extensive variables is that they scale with the number of atoms of the corresponding subsystem\cite{LL}. 
In particular, the average value of $\hat A_s$ is proportional to $N_s$,
\begin{equation}
\langle A_s \rangle \sim N_s ,
\end{equation}
and, therefore, $\langle \hat A_{1-M} \rangle \sim  N_1 + N_2 + \dots + N_M$. The key result is, now, that the variance of an extensive variable is also extensive. Let us see. Consider the variance of $\hat A_{1-M}$,
\begin{eqnarray}
\langle \hat A_{1-M}^2 \rangle - \langle \hat A_{1-M}\rangle^2 & = & \langle \sum_{j = 1}^M \sum_{k = 1}^M \hat A_j \hat A_k \rangle - \langle \sum_{j = 1}^M  \hat A_j \rangle \langle \sum_{k = 1}^M \hat A_k \rangle \nonumber \\
& = & \sum_{j = 1}^M \langle  \hat A_j^2 \rangle + \sum_{j \ne k} \langle \hat A_j \hat A_k \rangle - \sum_{j = 1}^M \langle  \hat A_j \rangle^2 - \sum_{j \ne k} \langle \hat A_j \rangle \langle \hat A_k \rangle .
\end{eqnarray}
We use now the fact that the subsystems are statistically independent to yield,
\begin{equation}
\langle \hat A_j \hat A_k \rangle = \langle \hat A_j \rangle \langle \hat A_k \rangle \>\>\>\>{\rm for}\>\>\>\> j \ne k.
\end{equation}
Hence, we obtain,
\begin{equation}
\langle \hat A_{1-M}^2 \rangle - \langle \hat A_{1-M}\rangle^2 = \sum_{j = 1}^M \left( \langle  \hat A_j^2 \rangle -  \langle  \hat A_j \rangle^2\right) .
\end{equation}
Namely, the variance of an extensive variable is also extensive and, therefore, it scales with the number of atoms in its subsystem,
\begin{equation}
 \langle  \hat A_s^2 \rangle -  \langle  \hat A_s \rangle^2 \sim N_s .
 \end{equation}
 As a matter of fact, all of the cumulants $\langle  (\hat A_s  - \langle   \hat A_s \rangle)^n \rangle$, with $n = 2, 3, 4, \dots$, are all extensive and, thus, {\it all} scale with $N_s$, i.e. $\langle  (\hat A_s  - \langle   \hat A_s \rangle)^n \rangle \sim N_s$.
 
Since the fluctuation, or width of the distribution ${\cal W}(A_s)$, is proportional to the square-root of the variance $\delta A_s = \left( \langle  \hat A_s^2 \rangle -  \langle  \hat A_s \rangle^2 \right)^{1/2}$, then this scales as $\delta A_s \sim \sqrt{N_s}$. On the other hand, as the mean value $\langle A_s \rangle$ is at or very near the peak of the distribution ${\cal W}(A_s)$, then, the result $\delta A_s/\langle A_s \rangle \sim 1/\sqrt{N_s}$ implies that ${\cal W}(A_s)$ is sharply peaked at the mean value, if $N_s \gg 1$. In other words, $\hat A_s$ essentially does not fluctuate and, effectively, looses its ``quantum" property of ``being" a superposition of more than one state: at every time, it is just in ``one state", its mean. This is not quite true, the system can still be in many different states, however, the density of states is so dense that, up to a precision $1/\sqrt{N_s}$, one cannot experimentally distinguish among them.  

One may believe that there are many sensible physical variables that are extensive. This is not the case. As a matter of fact, the only extensive mechanical variables that can be constants of motion in a closed system are the total energy,  momentum and angular momentum\cite{LL}. In a typical system studied from a thermodynamic point of view, both the angular and linear momenta can be taken as zero since the system is assumed to be at rest in the lab. These conditions give rise to the preponderance  of the internal energy in determining the states of systems in thermal equilibrium. Moreover, if the statistically independence of macroscopic subsystems holds true, this right away indicates why one can always associate well defined internal energies to macroscopic systems even if not in thermal equilibrium. One may say that the different subsystems ``measure" among themselves their internal energy at any time - barring exceptional states that one may prepare in the lab. At the same time, this shows why it is difficult to ``prepare" an otherwise closed system in a state that does not have its energy sharply peaked. With the previous ideas, the classical behavior of the motion of the center of mass follows very simply, as we now exhibit it.

The degrees of freedom of the little stone, denoted by $(\hat{\vec r}_1, \dots, \hat{\vec r}_N; \hat{\vec p}_1, \dots, \hat{\vec p}_N)$, can be separated into  center of mass and relative coordinates. These are,
\begin{eqnarray}
\vec R &=& \frac{1}{N} \sum_{i = 1}^N \vec r_i \nonumber \\
\vec {\bf r}_j & = & \vec r_j - \vec r_{j+1} \>\>\> j = 1, 2, \dots, N-1. \label{coordq}
\end{eqnarray}
The corresponding conjugated momenta are,
\begin{eqnarray}
\vec P &=& \sum_{i = 1}^N \vec p_i \nonumber \\
\vec {\bf p}_j & = & \sum_{k=1}^{N-1} K_{jk} \vec p_k \>\>\> j = 1, 2, \dots, N-1 , \label{coordp}
\end{eqnarray}
where the symmetric $(N-1) \times (N-1)$ matrix $K_{jk}$ is equal to,
\begin{equation}
K_{jk} = 2 \delta_{j,k} - \delta_{j,k-1} - \delta_{j,k+1} .
\end{equation}
With this, we can write the Hamiltonian as,
\begin{equation}
\hat H = \frac{\hat{\vec P}^2}{2 N m} + \hat H_{\rm rel} + \sum_{i=1}^N V_{\rm ext}\left(\hat {\vec R} + \sum_{k=1}^{N-1} a_k^{(i)} \hat {\vec {\bf r}}_k \right) \label{HCMREL}
\end{equation}
where $\hat H_{\rm rel}$ is the Hamiltonian of the relative coordinates $(\hat {\vec {\bf r}}_k,\hat {\vec {\bf p}}_k)$, $k = 1, 2, \dots, N-1$; it looks cumbersome and so we do not write it. The quantities $a_k^{(i)}$ are the coefficients of the inverse of the transformation of coordinates given by (\ref{coordq}), not explicitly written, but with the next two useful properties:
\begin{eqnarray}
\sum_{i=1}^N a_j^{(i)} & = & 0 \>\>\> \forall \>\>\>j = 1, 2, \dots, N-1 \nonumber \\
\sum_{i=1}^N a_j^{(i)}a_k^{(i)} & = & K_{jk}. \label{ides}
\end{eqnarray}
The relevant point we want to make is that the relative degrees of freedom are decoupled from the center of mass coordinates $(\hat{\vec R},\hat{\vec P})$, except for the external potential term, last in Eq.(\ref{HCMREL}). It is this last term the responsible for the ``measurement" of the center of mass position and momentum, within a very narrow latitude. 

Consider first two very common external potentials, one, the field produced by gravity, and other, a harmonic oscillator,
\begin{eqnarray}
V_{\rm ext}^{g}(\vec r) &=& m g z \nonumber\\
 V_{\rm ext}^{HO}(\vec r) &=& \frac{1}{2} m \omega^2 \vec r^2
 \end{eqnarray}
 with $z$ the coordinate in the direction of gravity. Note that  $m$ is the atomic mass. Using this in the last term of Eq.(\ref{HCMREL}) yields,
\begin{eqnarray}
\sum_{i=1}^N V_{\rm ext}^{g}\left(\hat {\vec R} + \sum_{k=1}^{N-1} a_k^{(i)} \hat {\vec {\bf r}}_k \right) & = & Nm g \hat Z \nonumber \\
\sum_{i=1}^N V_{\rm ext}^{HO}\left(\hat {\vec R} + \sum_{k=1}^{N-1} a_k^{(i)} \hat {\vec {\bf r}}_k \right) & = & \frac{1}{2} Nm \omega^2 {\hat {\vec R}^2} + \frac{1}{2} m \omega^2 \sum_{j=1}^{N-1} \sum_{k=1}^{N-1} K_{jk}  \hat {\vec {\bf r}}_j \cdot  \hat {\vec {\bf r}}_k , \label{easy}
\end{eqnarray}
with $Z$ the coordinate of $\vec R$ in the direction of gravity; to derive these expressions we used the identities (\ref{ides}). Note that in both cases the relative and center of mass motion are completely decoupled. It is also of interest to highlight  the known result that the mass associated to the center of mass coordinates is the total mass of the stone, $Nm$; it is of interest because it yields a ``macroscopic" potential for that coordinate. It is hard to imagine a macroscopic stone oscillating at frequencies $\omega$ of the order of megahertz or higher. So, we must consider, realistically, much smaller frequencies. This is in accord with the statement that the external potential must be macroscopic. From Eqs. (\ref{easy}) we find that, in order to couple the center of mass and relative degrees of freedom, the potential must have anharmonic terms. In the case of a stone falling in the field of gravity, the presence of the floor of the lab will do it: while the stone is falling, the relative motion does not feel gravity. However, during the collision of the stone with the floor, the relative degrees of freedom couple to the center of mass motion and the kinetic energy of the center of mass is transferred to the relative motion, heating the stone up.

Now, considering a generic anharmonic potential $V_{\rm ext}(\vec r)$, with an absolute minimum to make it simpler, and with spatial variations in macroscopic scales (i.e. with energy transitions much smaller than atomic ones), we can make an expansion around the center of mass coordinate $\hat {\vec R}$,
\begin{equation}
\sum_{i=1}^N V_{\rm ext}\left(\hat {\vec R} + \sum_{k=1}^{N-1} a_k^{(i)} \hat {\vec {\bf r}}_k \right)  \approx N V_{\rm ext}(\hat {\vec R}) + \frac{1}{2} \hat {\cal I}_{\alpha\beta}^{\rm (rel)} \frac{\partial^2}{\partial \hat R_\alpha \partial \hat R_\beta} V_{\rm ext}(\hat {\vec R})   \label{efepot}
\end{equation}
where $\alpha$ and $\beta$ are indices over the three spatial coordinates and summation is implied. The tensor $\hat {\cal I}_{\alpha\beta}^{\rm (rel)}$ operates on the relative variables only. It is given by,
\begin{equation}
\hat {\cal I}_{\alpha\beta}^{\rm (rel)} = \sum_{k=1}^{N-1} K_{jk}  \hat { {\bf r}}_{j\alpha}  \hat {{\bf r}}_{k\beta} \label{Iab}
\end{equation}
and it is related to the moment of inertia tensor of the body. The first term in (\ref{efepot}) is the external potential felt by the center of mass only and the second one is the coupling between relative and center of mass motion; clearly, $V_{\rm ext}(\vec R)$  must be anharmonic, otherwise there is no coupling. There may be situations, the collision of the stone with a floor mentioned previously as an example, where the above expansion may not be possible to perform. We shall assume the expansion is allowed just for purposes of exposition.

Let us now write Heisenberg equations of motion for the center of mass, $\gamma = x, y , z$,
\begin{eqnarray}
\frac{d}{dt} \hat{ R}_\gamma & = & \frac{\hat{ P}_\gamma}{Nm} \nonumber \\
\frac{d}{dt} \hat{P}_\gamma & = & - N \frac{\partial}{\partial \hat{ R}_\gamma} V_{\rm ext}(\hat{\vec R}) -
\frac{1}{2} \hat {\cal I}_{\alpha\beta}^{\rm (rel)} \frac{\partial}{\partial \hat{R}_\gamma} \frac{\partial^2}{\partial \hat R_\alpha \partial \hat R_\beta} V_{\rm ext}(\hat {\vec R}) .\label{Heis}
\end{eqnarray}
Similar equation can be written for the relative variables and those are coupled to the above ones through the last term in Eq. (\ref{efepot}).

Once again, for linear or quadratic external potentials, the last term in the second equation of (\ref{Heis}) vanishes and the motion of the center of mass becomes decoupled from the relative coordinates. In such a situation the density matrix of the system can be written exactly as,
\begin{equation}
\hat \rho(t) = \hat \rho_{\rm CM}(t) \hat \rho_{\rm rel}(t) ,\label{nocop}
\end{equation}
where $\hat \rho_{\rm CM}$ and $\hat \rho_{\rm rel}$ are the center of mass and relative degrees of freedom density matrices respectively. In this case, although the motion of the center of mass corresponds to a particle of macroscopic mass $Nm$, its dynamics is fully quantum and no classical limit can be obtained. 

However, if the center of mass and relative degrees of freedom are coupled, by the last term in the second equation of (\ref{Heis}), then, first of all, the full density matrix does not factorize as in Eq.(\ref{nocop}), i.e.,  $\hat \rho(t) \ne 
\hat \rho_{\rm CM}(t) \hat \rho_{\rm rel}(t)$. However, it is this coupling what allows to obtain
the classical limit of the center of mass motion, as we now argue. We recall, first, that both the center of mass position and momentum operators can be expressed in terms of the original positions and momenta of the $N$ particles of the sytem $(\hat {\vec r}_i,\hat {\vec p}_i)$, $i = 1, 2, \dots, N$, see Eqs. (\ref{coordq}) and (\ref{coordp}). The inverse transformation indicates that the original variables are functions of center of mass and relative variables. But since the latter are coupled, then, all variables become dependent among them. Now it enters the particularity of the model at hand, which is supposed to represent a little, solid stone. That is, because of the assumed short-range interactions, the original variables can depend on all of them only through collisions with their neighbors in the solid. This means that we can consider that the stone can be separated in even smaller pieces, yet macroscopic, such that the interactions between neighboring pieces  occur only at their borders. Under these conditions we assume, once more, that these little pieces become statistically independent.

To be a bit more precise, we can write the center of mass variables as,
\begin{eqnarray}
\vec R &=& \frac{1}{{\cal N}m} \sum_{s = 1}^{\cal N} {\cal M}_s \vec {\cal R}_s \nonumber \\
\vec P &=& \sum_{s = 1}^{\cal N} \vec {\cal P}_s \label{rel-small}
\end{eqnarray}
where $(\vec {\cal R}_s,\vec {\cal P}_s)$, $s = 1, 2, \dots, {\cal N}$, are the center of mass variables of the $s$-th piece. ${\cal M}_s$ are the masses of the pieces. Since the pieces are macroscopic, the number ${\cal N}$, while arbitrary, is not macroscopic and depends on the size of the stone. 

Due to all previous discussions, the variables of the centers of mass of the pieces are ``weakly coupled" and, therefore, typically, are statistically independent. At the end of the paper we discuss the meaning of ``typically". Under this result (or assumption) and because of the relations given by Eq. (\ref{rel-small}), the center of mass variables $(\hat {\vec R},\hat {\vec P})$ become extensive. As we have already shown, this implies that their fluctuations $\delta R_{\alpha}$ and $\delta P_\alpha$ scale as $N^{1/2}$ and, therefore, the (measurable) distribution of values of center of mass position and momentum are sharply peaked. 

Let us define the mean values of the center of mass variables as,
\begin{eqnarray}
\vec R(t) & = & \langle \hat{\vec R}(t) \rangle \nonumber \\
\vec P(t) & = & \langle \hat{\vec P}(t) \rangle
\end{eqnarray}
where the expectation value is taken with the full density matrix of the system $\hat \rho(t)$. Then, the equations of motion for the mean values are,
\begin{eqnarray}
\frac{d}{dt} R_\gamma & = & \frac{P_\gamma}{Nm} \nonumber \\
\frac{d}{dt} P_\gamma & = & - N \frac{\partial}{\partial R_\gamma} V_{\rm ext}({\vec R}) -
\frac{1}{2} \left< \hat {\cal I}_{\alpha\beta}^{\rm (rel)}(t) \right> \frac{\partial}{\partial R_\gamma} \frac{\partial^2}{\partial R_\alpha \partial R_\beta} V_{\rm ext}({\vec R}) .\label{class}
\end{eqnarray}
In writing this equation we have assumed, for simplicity, that at this level of approximation the center of mass and relative coordinates are also statistical independent. Although we expect this condition to be true in typical states, it is not really necessary to impose it but certainly makes the argument much easier to follow. However, what we claim is truly correct is that $\vec R$ is indeed extensive, in the sense described above, in order to  obtain the very important property, 
\begin{equation}
\langle V_{\rm ext}(\hat{\vec R}) \rangle \approx V_{\rm ext}(\langle \hat {\vec R}\rangle) .
\end{equation}

Equation (\ref{class}) is the main result of this note. It indicates that, indeed, the means of center of mass position and momentum obey Newton classical equations of motion. But since we can drop the adjective ``mean" because the fluctuation and higher cumulants are negligible, we simply say that the center of mass variables do obey Newton equations. While Eqs. (\ref{class}) look like  ``simple"  Ehrenfest equations of motion, we stress that their validity necessarily and explicitly require that the body is macroscopic. This fact is reflected in the last term of the second equation in (\ref{class}). This term, as we analyze below, is of a ``dissipative" nature in the sense that it is responsible for the transfer of energy between the center of mass and the relative degrees of freedom. Even if the statistical independence of center of mass and relative variables were not accurate, their coupling would result in a dissipative term for the center of mass. 

For purposes of exposition we have not insisted in the time evolution of the relative degrees of freedom. It is clear that we can make further assumptions for their study. For instance, one can assume that the atoms in the solid vibrate around their equilibrium positions. This would yield a further separation of relative variables into ``rigid body" rotations and relative small oscillations that, in general, would be coupled. In the appropriate limit, the rigid body degrees of freedom should also obey classical equations of motion, coupled to the center of mass and the relative vibrations. We believe that the main point of this note does not need to make this further analysis explicit.


To conclude the argument, we can now study the time evolution of the energy of the center of mass, defined by the operator,
\begin{equation}
\hat E_{\rm CM} = \frac{\hat{\vec P}^2}{2 N m} + NV_{\rm ext}(\hat {\vec R}) .
\end{equation}
By itself, it is the Hamiltonian of an ``isolated" particle of mass $Nm$ in the presence of the external potential $NV_{\rm ext}$. It is of interest to note that while it is fully quantum, it refers to a single macroscopic ``particle". The time evolution of $\hat E_{\rm CM}$ is given by,
\begin{eqnarray}
\frac{\partial }{\partial t} \hat E_{\rm CM} = \frac{i}{\hbar} \left[\hat H,\hat E_{\rm CM} \right]
\end{eqnarray}
with $\hat H$ the full Hamiltonian, given by Eq. (\ref{H0}). Using the assumed expansion of the external potential given by Eq. (\ref{efepot}), one finds, to the same order of approximation, 
\begin{equation}
\frac{\partial }{\partial t} \hat E_{\rm CM} \approx-  \frac{1}{2Nm} \hat {\cal I}_{\alpha\beta}^{\rm (rel)} \left( \frac{\partial}{\partial \hat R_\gamma} \frac{\partial^2}{\partial \hat R_\alpha \partial \hat R_\beta} V_{\rm ext}({\hat {\vec R}})\right) \hat P_\gamma .\label{Eclass}
\end{equation}
Defining $E_{\rm CM}(t) = {\rm Tr} \hat \rho(t) \hat E_{\rm CM}$, and using the statistical independence of the pieces of the stone, we obtain
\begin{equation}
\frac{\partial }{\partial t} E_{\rm CM}  \approx-  \frac{1}{2} \left< \hat {\cal I}_{\alpha\beta}^{\rm (rel)}(t) \right> \left( \frac{\partial}{\partial R_\gamma} \frac{\partial^2}{\partial R_\alpha \partial R_\beta} V_{\rm ext}({ \vec R})\right)  \frac{d R_\gamma}{dt} , \label{ecm}
\end{equation}
namely, a clear dissipative behavior that leads to an eventual mechanical equilibrium state where the energy of the center of mass of the stone will no longer change. It does not directly imply that the energy of the stone is dropping all the time from the initial state. There may exist intervals of time of decrease or of increase of energy, depending on the details of the initial condition and the potential, but given that there are a very large number of relative degrees of freedom, $N \gg 1$, we expect from general considerations of the irreversible behavior of macroscopic bodies, that eventually the energy $E_{\rm CM}$ of the stone will tend to a minimum. In this situation, all the extra initial energy of the stone will be released to the relative degrees of freedom. The stone, in turn, will end up with a higher temperature than initially. It is also interesting to observe that when we do classical mechanics theory for the motion of a stone, we usually neglect the dissipative term in Eqs. (\ref{class}) and (\ref{Eclass}), thereby assuming that the motion of the center of mass in the external potential is conservative. While we do know that this is never true in real life, we find from the present argument, a glimpse of how the dissipation is always there and, furthermore, that its origin can be traced back to quantum mechanics.

The present ``derivation" or exposition of requirements, for the attainment of classical motion of the center of mass of a macroscopic stone, has certainly hidden the problems of ``decoherence" and ``measurement", typical of all discussions of QM when applied to macroscopic bodies\cite{Zurek}. To begin, we have avoided to discuss the processes, and the time they take, that lead an arbitrary initial state to one in which statistical independence has been reached; this is the problem of decoherence. However, we claim, ``typical" initial conditions  in the lab, for a real stone of 10$^{20}$ atoms are already in states where statistical independence is essentially valid. One would have to work very hard to devise a procedure to prepare a stone in an initial state, say, in which its energy was not already sharply peaked. It is important to observe that the part of the system composed of the relative degrees of freedom can be at very low temperatures, such that a quantum many-body treatment would still be required. That is, the statement of the emergence of classical behavior refers to the center of mass degrees of freedom only. Yet, we do admit, we have not shown how decoherence actually occurs. The second issue, that of measurement, we have already alluded to it. That is, once we accept that statistical independence of the macroscopic little pieces that conform the stone sets in, these pieces ``measure" among themselves all their extensive variables, giving them sharp values at all times.  And we recall once more, the crucial role played by the macroscopic external potential in determining the extensive statistical property of the center of mass variables. But again, this nothing but the essence of the success of the theory of statistical physics. 

Acknowledgment is given to UNAM IN107014 and CONACYT LN-232652 for support.


\begin{thebibliography}{99}

\bibitem{Cohen} C. Cohen-Tannoudji, B. Diu, and F. Laloe, {\it Quantum Mechanics I and II}, Wiley, New York, 1977.

\bibitem{Zurek} W.H. Zurek, {\it Decoherence and the transition from quantum to classical - revisited}, Los Alamos Science, Number 27, 2002; and references therein.

\bibitem{LL} L. Landau and L. Lifshitz, {\it Statistical Physics I and II}, Pergamon, Oxford, 1980. 

\end{thebibliography}
\end{document}